\documentclass[pra,aps,twocolumn,floatfix,superscriptaddress]{revtex4-2}
\usepackage{epsfig} 
\usepackage{graphicx} 
\usepackage{color}  
\usepackage{hyperref}
\usepackage{bibentry}

\usepackage{dsfont}
\usepackage{amsmath}
\usepackage{amsfonts}
\usepackage{amssymb}
\usepackage{bbm}
\usepackage{bm}
\usepackage[normalem]{ulem}

\makeatletter
\newcommand*{\balancecolsandclearpage}{%
  \close@column@grid
  \clearpage
  \twocolumngrid
}
\makeatother

\makeatletter
\let\cat@comma@active\@empty
\makeatother

\newcommand{\br}{{\mathbf r}}

\definecolor{darkgreen}{rgb}{0.0, 0.5, 0.0}

\newcommand{\potContDist}{{u_\mathrm{d}}}
\newcommand{\potContTilt}{{u_\mathrm{t}}}
\newcommand{\potContDistFinal}{{u^{\star}_\mathrm{d}}}
\newcommand{\dimlessZ}{{\bar{z}}}

\begin{document} 
\title{Relaxation in an Extended Bosonic Josephson Junction}

\author{J.-F.~Mennemann}
\affiliation{Research Platform MMM ``Mathematics--Magnetism--Materials", c/o Fak.~f{\"ur} Mathematik, Univ.~Wien, 1090 Vienna, Austria}
\affiliation{Wolfgang Pauli Institute c/o Fak.~f\"{u}r Mathematik, Univ.~Wien, Oskar-Morgenstern-Platz 1, 1090 Vienna, Austria}

\author{I.~E.~Mazets}
\affiliation{Wolfgang Pauli Institute c/o Fak.~f\"{u}r Mathematik, Univ.~Wien, Oskar-Morgenstern-Platz 1, 1090 Vienna, Austria}
\affiliation{Vienna Center for Quantum Science and Technology, Atominstitut, TU~Wien,~Stadionallee~2,~1020~Vienna,~Austria}

\author{M.~Pigneur}
\affiliation{Vienna Center for Quantum Science and Technology, Atominstitut, TU~Wien,~Stadionallee~2,~1020~Vienna,~Austria}

\author{H.~P.~Stimming}
\affiliation{Research Platform MMM ``Mathematics--Magnetism--Materials", c/o Fak.~f{\"ur} Mathematik, Univ.~Wien, 1090 Vienna, Austria}

\author{N.~J.~Mauser}
\affiliation{Research Platform MMM ``Mathematics--Magnetism--Materials", c/o Fak.~f{\"ur} Mathematik, Univ.~Wien, 1090 Vienna, Austria}

\author{J.~Schmiedmayer}
\affiliation{Vienna Center for Quantum Science and Technology, Atominstitut, TU~Wien,~Stadionallee~2,~1020~Vienna,~Austria}

\author{S.~Erne}\email{erne@atomchip.org}
\affiliation{Vienna Center for Quantum Science and Technology, Atominstitut, TU~Wien,~Stadionallee~2,~1020~Vienna,~Austria}

\date{\today}

\begin{abstract} 
We present a detailed analysis of the relaxation dynamics in an extended bosonic Josephson junction. We show that stochastic classical field simulations using Gross-Pitaevskii equations in three spatial dimensions reproduce the main experimental findings of M.~Pigneur \textit{et al.}, Phys.~Rev.~Lett.~\textbf{120}, 173601 (2018). We give an analytic solution describing the short time evolution through multimode dephasing. For longer times, the observed relaxation to a phase locked state is caused by nonlinear dynamics beyond the sine-Gordon model, induced by the longitudinal confinement potential and persisting even at zero temperature. Finally, we analyze different experimentally relevant trapping geometries to mitigate these effects. Our results provide the basis for future experimental implementations aiming to study nonlinear and quantum effects of the relaxation in extended bosonic Josephson junctions.
\end{abstract}

\maketitle 

\section{Introduction}

The Josephson effect is a prominent example for the manifestation of macroscopic quantum effects. Originally formulated in the context of two weakly coupled superconductors \cite{JOSEPHSON1962251} it shows a broad range of applications for systems featuring two coupled macroscopic quantum states. As such Josephson junctions were observed and analyzed in a variety of systems (see e.g.~\cite{barone1982physics}). Increased interest over the last decades has been on its application to atomic systems, where two-body interactions enrich the dynamical behavior. This has led to a number of ongoing theoretical and experimental studies, from fermionic superfluids \cite{Valtolina1505}, macroscopic quantum self trapping \cite{Raghavan99,Albiez2005}, bosonic Josephson junctions \cite{Javanainen86,Jack96,Smerzi97,levy2007ac,gati2007bosonic,leblanc2011dynamics}, to different geometries \cite{whitlock2003relative,Likharev79} from small to extended junction arrays.

\begin{figure*}[t!]
\centering
\includegraphics[width=0.9\textwidth]{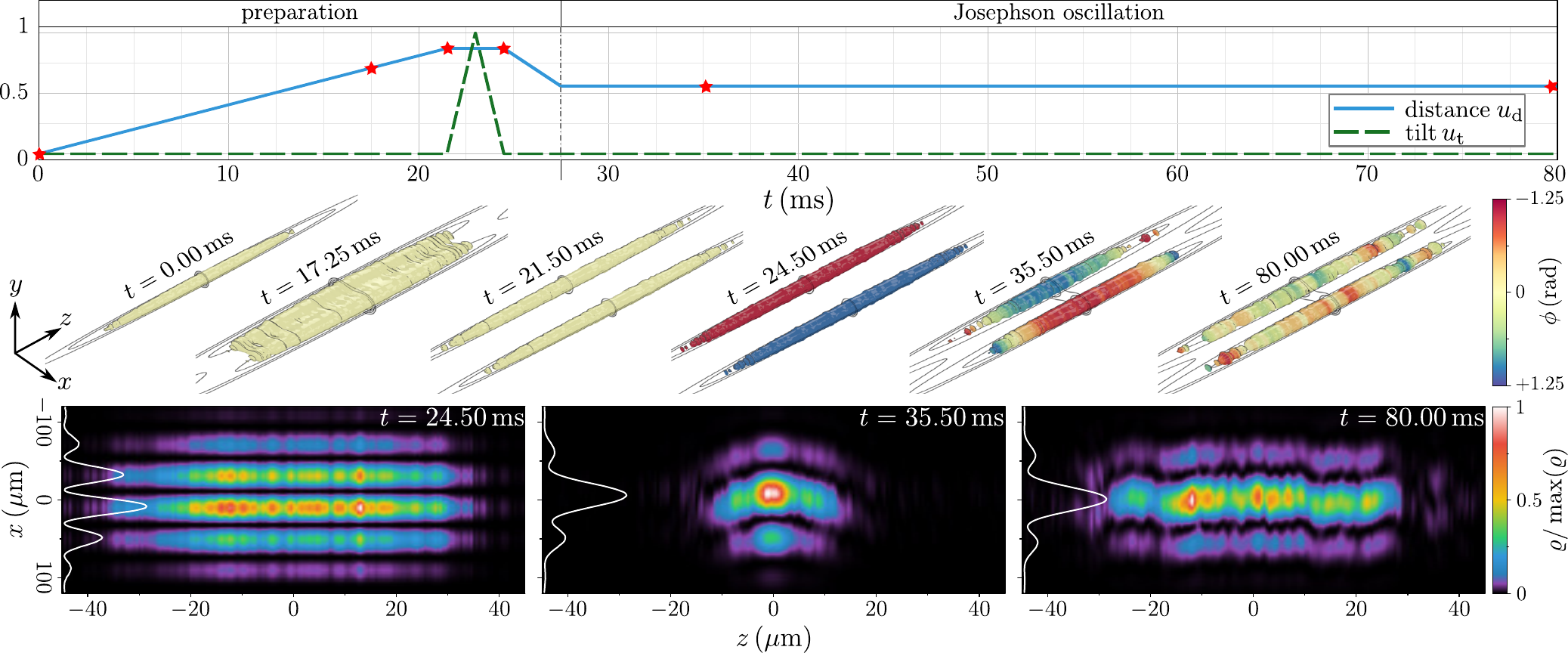}
\caption{
Schematic of the experimental sequence \cite{Pigneur18}. 
Upper panel: 
During preparation a single condensate is coherently split (blue line shows the normalized dressing amplitude $\potContDist(t)$). 
Once completely decoupled, a tilt (dashed green line shows the normalized tilt factor $\potContTilt(t)$) 
along the transversal $x$-coordinate 
introduces an energy difference between the wells resulting in the accumulation of a global 
phase difference $\Phi_0$. 
Thereafter, a variable coupling is achieved by reducing the distance on a short timescale, 
leading to the observed Josephson oscillations. 
Middle panel: Snapshots of the wavefunction at different times (red stars from left to right) for a single run that samples thermal initial conditions for temperature $T \approx 20$\,nK with $N=3500$ particles and $\Phi_0=-1.25\,\mathrm{rad}$.
The 3D figures show isosurfaces of the density ($0.35$ of the maximum value) with the local 
phase difference $\phi(\br,t)$ encoded in color. 
Grey lines show isolines of the external potential $V(\br,\potContDist, \potContTilt)$. 
Lower panel: Interference pictures after finite time of flight with the normalized density 
depicted in color and the integrated (along the $z$-direction) transverse density depicted in white. 
The almost straight interference fringes at $t=80\,\mathrm{ms}$ show relaxation to a phase locked state.
}
\label{fig_2}
\end{figure*}

A recent experiment \cite{Pigneur18}, studying an extended bosonic Josephson junction in an extended one-dimensional (1D) geometry with two elongated (quasi-1D) $^{87}$Rb superfluids in a double-well potential (see Fig.~\ref{fig_1} and Fig.~\ref{fig_2}), makes the situation even less trivial. The observed Josephson oscillations (of the interwell atom-number difference or of its conjugate variable, the global phase difference) were damped after only a few periods. More intriguing, this relaxation led to a \textit{phase locked state} with strongly reduced fluctuations, observed experimentally through the almost straight interference fringes  along the longitudinal axis of the trap (see Sec.~\ref{sec:2} for details). This implies local damping of Josephson oscillations within each experimental realization and relaxation beyond simple dephasing dynamics. To date, this behavior could not be explained by various microscopic models \cite{whitlock2003relative,van2020low,van2020josephson} or its low energy effective description, the sine-Gordon model, not only in the semiclassical, but also in the quantum \cite{Torre13,Foini17,van2019self,horvath2019nonequilibrium} treatment. 

Here we present a detailed numerical study of the quasicondensate dynamics, which explains the main results of the experiment \cite{Pigneur18}. For the experimentally relevant harmonic confinement we show that the system relaxes in two stages. The short time dynamics is fully described by multimode dephasing, already leading to a local damping of the oscillations. At longer times, we show that nonlinear dynamics beyond the sine-Gordon model causes the relaxation of the system to the observed phase locked state. We find this effect to persist even at zero temperature, which highlights the importance of understanding the relevance of classical nonlinear dynamics of thermally fluctuating fields when analyzing complex quantum many-body systems. Finally, we discuss different experimentally realizable trapping geometries to mitigate these effects.

The article is organized as follows. We start in Sec.~\ref{sec:2} with an overview of our numerical simulations for the experimental system considered in \cite{Pigneur18} and give details of the numerical implementation and calculation of experimentally relevant observables. In Sec.~\ref{sec:3} we present numerical results, reproducing the main findings of \cite{Pigneur18} over a wide range of initial conditions, and discuss the observed relaxation to a phase locked state. Lastly, in Sec.~\ref{sec:4}, we compare our numerical results to the effective one-dimensional description of the system, determining the microscopic origin of the observed relaxation and possibilities to mitigate the effect. We conclude our work in Sec.~\ref{sec:5}.

\section{Numerical model and observables} \label{sec:2}

\begin{figure}[t!]
\centering
\includegraphics[width=0.455\textwidth]{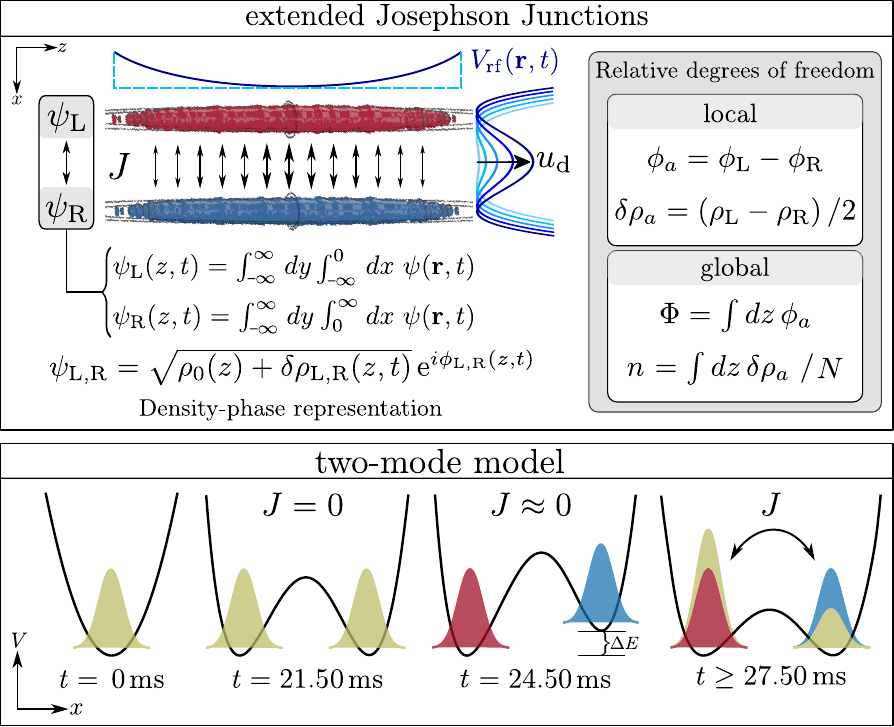}
\caption{
Schematic of the extended bosonic Josephson junction of \cite{Pigneur18}. We consider two tunnel-coupled superfluids (red and blue ellipses) in a double-well (DW) potential (blue lines). By adjusting the barrier height the tunneling coupling $J$ can be adjusted. The superfluids $\psi_\mathrm{L,R}$ are spatially separated and can be described in terms of density fluctuations $\delta \rho_\mathrm{L,R}$ around a mean density profile $\rho_0$ and a fluctuating phase $\phi_\mathrm{L,R}$ (depicted in color). The tunneling-coupling $J$ influences the antisymmetric phase $\phi_a$ and density fluctuations $\delta \rho_a$. Considering only the global phase $\Phi$ and conjugate density difference $n$ reduces the system to a two-mode model. The lower panel depicts schematically the experimental sequence (see also Fig.~\ref{fig_2}) from splitting a single condensate, introducing a phase difference (depicted in color) by applying a small tilt $\Delta E$ to the decoupled DW, to the Josephson oscillation regime where the finite tunneling-coupling $J$ leads to coherent oscillations of particles between the two wells.
}
\label{fig_1}
\end{figure}

Our numerical simulation follows closely the experiment in \cite{Pigneur18}
where the extended bosonic Josephson junction was realized through an ultracold gas of $^{87}$Rb atoms 
in an adjustable double-well potential on an atom chip. 
The gas was cooled well below degeneracy, such that the evolution of the system can be described 
by means of the Gross--Pitaevskii equation (GPE)
\begin{equation}
\label{eq:gpe}
\begin{aligned}
i \hbar \frac \partial {\partial t} \Psi (\br, t) 
&=
\Big[
-\frac {\hbar ^2}{2m} \nabla^2 + V(\br,t) \\
&\qquad + \frac {4\pi \hbar ^2a_s} m |\Psi (\br,t)|^2 
\Big] \Psi (\br,t)\,.
\end{aligned}
\end{equation} 
The GPE describes the evolution for the order parameter $\Psi(\br,t)$ of a quasicondensate \cite{pitaevskii2016bose}, capturing the contribution of nonlinear dynamics of classical fields. 
Here $m$ and $a_s$ are the atom mass and $s$-wave scattering length of $^{87}$Rb, respectively, and $V$ is the external confinement potential modeled as
\begin{equation}
\label{eq:V}
V\big( \br,\potContDist(t),\potContTilt(t) \big) 
= V_\mathrm{rf}\big( \br, \potContDist(t) \big) - \potContTilt(t) x ~.
\end{equation}
The adiabatic radio-frequency potential $V_\mathrm{rf}(\br, \potContDist)$ on an atom chip~\cite{Lesanovsky06} has a weak confinement along the longitudinal $z$-direction and can be continuously deformed from a single-well to a double-well (DW) potential (along the $x$-direction) for increasing values of $\potContDist$ (see Fig.~\ref{fig_1}). The control parameter $\potContDist$, also known as the (normalized) dressing amplitude, determines the distance between the two wells, i.e.~the height of the potential barrier in the DW. By means of the second control parameter $\potContTilt$ it is possible to apply a small energy difference with respect to the transverse $x$-coordinate, leading to a tilted DW configuration used to initialize the Josephson oscillations in the experiment. 

Due to the tight radial confinement within each well, $\nu_\perp \gg \nu_\parallel$, the system consists of two spatially separated elongated superfluids (see Fig.~\ref{fig_1}). If the typical energy scales of the gas are small compared to the energy of the first radially excited state, i.e.~$\mu, k_\mathrm{B} T \ll \hbar \omega_\perp$, the system is in the quasi-1D regime, with the dynamics along the radial directions effectively frozen. Note that atomic repulsion may lead to a broadening of the radial wave function, which can be taken into account in the adiabatic limit \cite{salasnich2002effective}. We define the one-dimensional antisymmetric (relative) phase
\begin{align}
\phi_a(z,t) &= \phi_\mathrm{L}(z,t) - \phi_\mathrm{R}(z,t) ~,
\end{align}
and conjugate 1D density difference
\begin{align}
\delta \rho_a(z,t) &= \Big( \rho_\mathrm{L}(z,t) - \rho_\mathrm{R}(z,t) \Big) / 2 ~,
\end{align}
with $\phi_\mathrm{L,R}$ and $\rho_\mathrm{L,R}$ the longitudinal 1D phase and density profile of the left and right component, respectively (c.f.~Fig.~\ref{fig_1}). In the following we drop the subscript `\textit{a}' when there is no risk of confusion. Note that, the potential in~\eqref{eq:V} is non-separable, which requires additional approximations when reducing the effective dimension of the system.

Josephson effects in bosonic systems lead to a coherent oscillation of particles between the two sides of the DW via tunneling of atoms through the potential barrier (see Fig.~\ref{fig_1}). In the commonly used two-mode approximation the time evolution of the system is described by the Josephson equations (see e.g.~\cite{Javanainen86,Jack96,Raghavan99})
\begin{align}
\dot{n}(t) &\approx -2J \sqrt{1-n^2(t)} \operatorname{sin}(\Phi(t)) \label{eq:JJ_density}\\
\dot{\Phi}(t) &\approx \frac{\mu}{\hbar} n(t) + 2J \frac{n(t)}{\sqrt{1-n^2(t)}} \operatorname{cos}(\Phi(t)) \label{eq:JJ_phase} ~,
\end{align}
where $n(t) = (N_\mathrm{L}(t) - N_\mathrm{R}(t))/N$ and $\Phi(t) = \Phi_\mathrm{L}(t) - \Phi_\mathrm{R}(t)$ are the normalized atom number difference and conjugate relative phase difference between the two wells, respectively. Here $\mu$ is the on-site interaction energy and $2 \hbar J$ is the single particle tunneling coupling energy. In the following, we consider the Josephson regime $1 \ll \mu/(2 \hbar J) \ll N^2$. For small initial amplitudes the Josephson equations describe oscillations of the phase and particle imbalance characterized by the plasma, or Josephson, frequency $\omega_{J} \approx \sqrt{4 J \mu / \hbar}$. 

For extended bosonic JJ the applicability of the two-mode model strongly depends on the geometry of the system. In strongly elongated systems dynamics within each condensate may no longer be negligible, due to the high density of states along the extended direction. This naturally leads to a description of the system in the language of quantum field theory, where the Josephson equations \eqref{eq:JJ_density} and \eqref{eq:JJ_phase} are replaced by the sine-Gordon model (see e.g.~\cite{Gritsev07}) for the local fields $\phi(z,t)$ and $\delta \rho(z,t)$. A prominent example for the change in dynamical behavior is the breakdown of quantum self trapping (see e.g.~\cite{Hipolito2010}). Additionally, radially excited states may contribute to the dynamics, driving the system beyond the 1D regime. In order to reproduce the main findings of the original experiment we therefore decided to consider a full three-dimensional classical fields simulation.

In the following, we first give an overview of the experimental sequence before discussing in the remainder of this section details of the numerical implementation and calculation of observables.

\subsection{Overview of experimental sequence}

We depict in Fig.~\ref{fig_2} a typical evolution of the experimental system according to our numerical simulations. The experiment starts with a single quasi one-dimensional condensate in thermal equilibrium. The initial trap at $t=0$ has a weak harmonic confinement with frequency $\nu_\parallel = 22 \, \mathrm{Hz}$ along the longitudinal axis. The radial trapping frequencies are both given by $\nu_\perp \approx 3 \, \mathrm{kHz}$, such that the system is in the quasi-1D regime. 

The trap begins to split at $t>0$ and approaches at $t = 21.5 \, \mathrm{ms}$ its double-well configuration corresponding to the largest distance of the two halves of the quasicondensate. At this point the barrier of the DW is sufficiently high, such that the two 1D-BECs are completely decoupled. Unlike \cite{Gring1318,Langen207}, splitting is assumed to be rather close to the adiabatic limit \cite{Leggett98}, and we end up in the situation when the quantum fluctuations are negligible to thermal ones. Therefore, the phase and density fluctuations after the splitting were found to be small in the experiment \cite{Pigneur18}, signaling strong atom-number squeezing. This enables us to use finite-temperature classical statistical numerical methods, which do not include the quantum shot-noise, to describe the full experimental splitting process.

Next, a tilt along the transverse direction of the trap induces an energy difference between the two wells, which leads to the accumulation of a global phase difference $\Phi_0$ (see Fig.~\ref{fig_2} for $t=24.5 \, \mathrm{ms}$). The final phase difference $\Phi_0$ can be adjusted by altering the maximum value of the tilt factor $\potContTilt$. Note that no particle number difference accumulates due to the complete decoupling of the two clouds. Experimentally the local relative phase $\phi$ can be extracted from the interference pattern resulting from the overlap of the wavefunctions after finite time-of-flight. Numerical results for the interference patterns after free expansion are depicted in the lower panel of Fig.~\ref{fig_2}. In accordance with the experiment, the straight interference fringes at $t=24.5 \, \mathrm{ms}$ signal a well defined initial relative phase with negligible fluctuations along the whole condensate. This global relative phase $\Phi(t)$ can be extracted directly from the interference pictures by integrating along the longitudinal direction (white lines).

Lastly, within the period of $t = 24.5 \, \mathrm{ms}$ and $t = 27.5 \, \mathrm{ms}$ the distance of the two halves of the condensate is lowered to its final value ($\potContDist=\potContDistFinal$ and $\potContTilt=0$), decreasing the barrier height. Tunneling through the barrier couples the two wells and the system starts to oscillate, realizing the extended bosonic Josephson junction in the relative phase $\phi$ and conjugate density difference $\delta \rho$. The final potential for $t \geq 27.5\,\mathrm{ms}$ has a reduced harmonic confinement of $\nu_\parallel = 12 \, \mathrm{Hz}$ along the longitudinal direction and, within each well, an approximately harmonic confinement of $\nu_\perp \approx 1.5 \, \mathrm{kHz}$ along the radial direction. Both frequencies $\nu_\parallel$ and $\nu_\perp$ are in nice agreement with the parameters of the original experiment~\cite{Pigneur18}.

In accordance with the experiment the global Josephson oscillation is rapidly damped and the system relaxes to a quasi-stationary state with a global relative phase $\Phi \approx 0$ and strongly reduced phase fluctuations. This so called phase-locked state is apparent in the stationary, almost straight interference fringes observed at long times (see last interference pattern in Fig.~\ref{fig_2}) and signals relaxation beyond local dephasing. 

\subsection{Numerical implementation} \label{sec:2_2}

In order to prepare finite temperature initial conditions
for $\Psi(\br,0)$ we first compute ground state solutions of the GPE
using imaginary time propagation \cite{Chiofalo2000}.
In this context every wave function is normalized to a desired atom number $N$. 
The distribution of the total atom number $N$ is assumed to be a normal distribution 
$
f(N) = \mathcal{N}(\bar{N}, \sigma_N^2)
$
with the mean $\bar{N}=3500$ and standard deviation $\sigma_N=0.16\, \bar{N}$.
In the original experiment post\-selection further restricts the atom numbers
to the range $[\bar{N} - \delta_N, \bar{N} + \delta_N]$ with a given cut-off parameter $\delta_N=0.08\, \bar N$.
In the numerical simulation the restricted atom number distribution is obtained by inversely sampling
the cumulative distribution function $F(N)$
over the interval 
$[F(\bar{N} - \delta_N), F(\bar{N} + \delta_N)]$ using an equidistant 
distribution of $n_\mathrm{sr}=301$ points representing the number of single runs.

Subsequently, the resulting zero temperature ground state solutions are propagated using the 
stochastic Gross-Pitaevskii equation (SGPE) (see e.g.~\cite{blakie2008dynamics} and references therein)
\begin{equation}
\label{eq:sgpe}
\begin{aligned}
i \hbar \frac{\partial}{\partial t} & \Psi(\br,t)
=
(1 - i \gamma) 
\Big[
-\frac{\hbar^2}{2 m} \nabla^2  + V(\br, 0) \\
&+ \frac {4\pi \hbar ^2a_s} m |\Psi(\br, t)|^2 
-\mu
\Big] \Psi(\br, t)
+ \eta(\br, t)
\end{aligned}
\end{equation}
until a new stationary thermal state is reached.
Here, $\mu$ denotes the chemical potential of the eigenvalue problem at zero temperature and
$\eta$ is a complex random noise term.
For the simple growth SGPE considered here, the positive constant $\gamma$ can be freely tuned to improve the 
speed of convergence. 
We keep the atom number fixed within each SGPE realization by normalizing the wave function after each time step. 
This is done to preserve the exact atom number distribution already included in the above 
calculation of the ground state for each realization. 
Including the fluctuations of the SGPE would, however, only lead to a small broadening of the total atom number distribution. 

The numerical propagation of the SGPE is based on a second-order accurate operator splitting
and the spatial derivatives are approximated by means of the Fourier spectral collocation method.
We note that the thermal noise term $\eta$ is assumed to be constant for the duration of every time step.
Depending on the desired temperature, several tens of thousands of time steps are necessary to reach a thermal state.
In the simplest approximation, $\eta$ denotes a complex, Gaussian, white noise process with
correlations
\begin{equation}
\langle \eta^*(\br,t) \eta(\br', t') \rangle 
= 
2 \hbar \gamma k_\mathrm{B} T \delta(\br-\br') \delta(t-t')
\label{eq:eta_original}
\end{equation}
corresponding to a given temperature $T$.
However, using the noise term in form of Eq.~\eqref{eq:eta_original} results in unrealistic excitations 
of the quasicondensate along the tightly confined transverse directions of the trap.
One obvious solution to this problem is to project the wave function onto a few of the lowest energy 
single particle eigenstates of the harmonic trap.
This approach, which is known as the projected stochastic Gross-Pitaevskii 
equation \cite{blakie2008dynamics}, is prohibitively expensive in our three-dimensional setting. 
Due to the extremely strong transverse confinement of the initial trap $k_\mathrm{B} T \ll \hbar \omega_\perp$, 
the main effect of radially excited single particle states is an increase in width of 
the Gaussian ground state wave function \cite{salasnich2002effective}. We therefore expect the desired thermal 
state to be an almost perfectly symmetric and smooth function with respect to the transverse directions $x$ and $y$.

This assumption can be taken into account in the preparation of the thermal noise
$\eta(\br, t)$ in Eq.~\eqref{eq:sgpe}.
In this context, we first compute a complex field $\Psi^\perp(x,y,z) = \Psi(x,y,z) / \sqrt{\rho(z)}$
using the 1D density
\begin{align}
\rho(z) = \int \int |\Psi(x,y,z)|^2 \,dx\,dy
\end{align}
along the $z$-direction.
For the noise term we finally employ the expression
\begin{align}
\eta(x, y, z) = \lambda(z) \Psi^\perp(x,y,z) \, ,
\end{align}
where $\lambda(z)$ is one-dimensional Gaussian white noise with zero mean and variance 
given by Eq.~\eqref{eq:eta_original} replacing $\delta(\br-\br')$ with $\delta(z-z')$. In the continuum limit an explicit cutoff for the noise is necessary due to the Raleigh-Jeans divergence, leading to a smeared out delta function. We checked independence of the cutoff for selected parameters. Convergence to the correct thermal state was verified by further evolving the system with the GPE after slightly disturbing the exact symmetry of the prepared states in the $x,y$-plane.

Once a set of thermal initial conditions $\Psi(\br, 0)$ has been prepared,
the actual time evolution of all corresponding single runs is computed using the ordinary GPE~\eqref{eq:gpe}.
Similarly to the case of the SGPE we employ a second-order accurate operator splitting (Strang splitting)
in combination with a Fourier spectral discretization of the spatial 
derivatives~\cite{BaoJakschMarkowich03, BaoJinMarkowich03}.
It is worth noting that the number of atoms $N$ in every single run is conserved throughout the whole
simulation as it is expected from the given numerical algorithm.
We also checked that the total energy $E$ is conserved (with high precision) 
as long as the two external parameters $\potContTilt$ and $\potContDist$ remain 
unchanged (i.e.~the Hamiltonian is time-independent).

The computational effort of a classical fields simulation in three spatial dimensions
using several hundred initial states and on the order of $100\,000$ time steps is significant.
However, the algorithm can be nicely implemented on a graphics processing unit (GPU)
resulting in a dramatic speed-up in comparison with a CPU-based implementation 
on a multi-core CPU system.

\subsection{Experimental observables}

In the experiment, the evolution of the system is investigated by destructive measurements 
after time-of-flight (TOF) either looking at the interference pictures or the atom number difference 
between the two wells. We calculate the TOF expansion numerically, taking atom interactions into account 
for the first millisecond of the expansion. 
During this time, the system rapidly expands along the tightly confined radial direction, 
diluting the gas sufficiently such that the expansion becomes ballistic (i.e. non-interacting).
Once the TOF-expansion has been computed, we integrate the density along the vertical 
direction $y$ to obtain the desired interference pictures measured in the experiment.
Additionally, to account for the finite imaging resolution in the experiment, we include a 
convolution of the numerical interference pictures with a Gaussian point-spread function
\begin{align}
\xi(x,z) = \exp( -(x^2 + z^2) / (2 \sigma_\mathrm{psf}^2) ) / (2 \pi \sigma_\mathrm{psf}^2) ~.
\end{align}
The local relative phase $\phi(z,t)$ can be extracted by fitting a sinusoidal function to each pixel along the longitudinal $z$-direction. 
Finally, integrating along the longitudinal $z$-direction to obtain the profile $n_\mathrm{p}(x)$, 
the contrast $C_\mathrm{tof}$ and the global relative phase $\Phi_\mathrm{tof}$ are determined
using
\begin{equation}
n_\mathrm{p}(x) \approx A \exp(-x^2 / (2 \sigma^2)) \big( 1 + C_\mathrm{tof} \cos(k x - \Phi_\mathrm{tof}) \big),
\label{eq:profile_tof}
\end{equation}
where $A$, $\sigma$, $C_\mathrm{tof}$, $k$ and $\Phi_\mathrm{tof}$ are found by solving 
a nonlinear least squares problem. The contrast $C_\mathrm{tof}$ measures the visibility of the integrated interference fringes (c.f.~Fig.\ref{fig_2}), i.e. it takes its maximum value $C_\mathrm{tof} =1$ for negligible fluctuations of the relative phase along the whole condensate.

The computation of the global relative phase $\Phi_\mathrm{tof}$ by means of a TOF-simulation 
and formula~\eqref{eq:profile_tof} is a time-consuming process.
Alternatively, the global relative phase can be extracted from the in situ wave function (see e.g.~\cite{Imambekov2009,Nieuwkerk2018}) directly via
\begin{align}
&\Phi
= \arg \! 
\bigg[ 
\int_0^\infty 
\!\!\!\! dx\!\! 
\int_{-\infty}^\infty 
\!\!\!\! dy \!\! 
\int_{-\infty}^\infty 
\!\!\!\! dz 
~ \Psi(x,y,z) \Psi^*(-x,y,z) 
\bigg]. 
\label{eq:global_relative_phase}
\end{align}
The results of both methods are practically indistinguishable and the weighting implicitly applied in Eq.~\eqref{eq:global_relative_phase} reflects the weighting involved in the procedure using the TOF-images reasonably well.

Consistently, the local phase profile $\phi(z,t)$ can be defined by omitting the integration over $z$ in Eq.~\eqref{eq:global_relative_phase}. Due to coherence along the tightly confined radial direction, the results are reasonably close to the line values
\begin{align}
\phi(z) 
= 
\arg
\left[
\Psi(x_{1},0,z)\Psi^*(x_{2},0,z)
\right].
\label{eq:relative_phase_z_x1_x2}
\end{align}
Here, $x_1$ and $x_2$ are at the minimum of the radial potential in the left 
and right well (i.e. at the points of maximum density), respectively.
Equivalently, the contrast $C$ can be calculated from the in situ phase profiles via
\begin{align}
C
= 
\operatorname{Re} 
\left[ 
\int \! dz \int \! dz' \langle \operatorname{e}^{-i \left( \phi(z)-\phi(z') \right)} 
\rangle \right] \, ,
\end{align} 
where we neglected the strongly suppressed density fluctuations \cite{gritsev2006full}. We therefore consider in the following the in situ observables, omitting the time of flight expansion.

\section{Numerical results and analytic estimates for the experiment} \label{sec:3}

\begin{figure}
\centering
\includegraphics[width=0.48\textwidth]{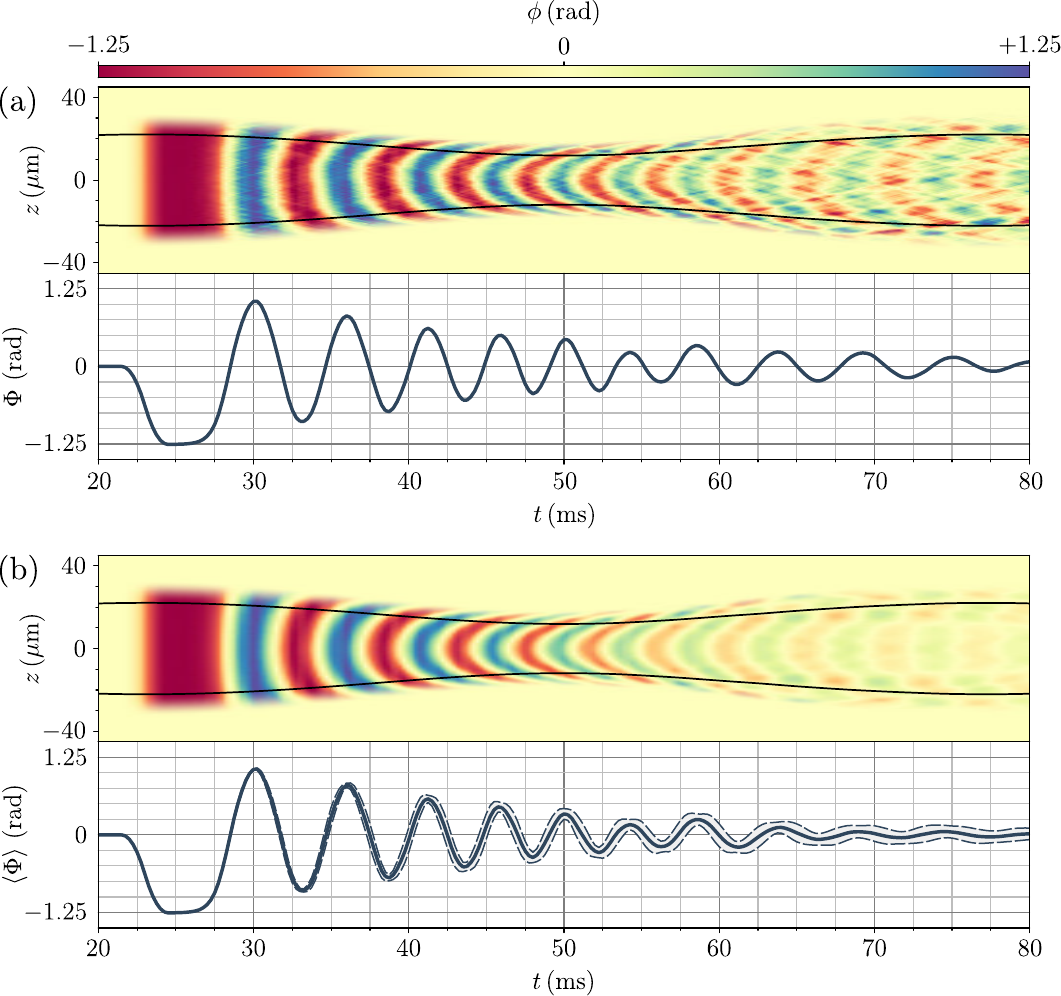}
\caption{
Results of the numerical simulation outlined in Fig.~\ref{fig_2}.
(a) Local relative phase $\phi(z,t)$ (upper panel) and global relative phase $\Phi(t)$ (lower panel) of a selected single run using $N=3500$ particles.
(b) Ensemble average of the local relative phase $\langle \phi(z,t) \rangle$ (upper panel)
and global relative phase $\langle \Phi(t) \rangle$ using $301$ realizations including experimental fluctuations of the total number of atoms. Dashed lines depict the standard deviation.
Black lines in (a) and (b) mark the region containing $75 \%$ of the atoms.
}
\label{fig_3}
\end{figure}

Numerical results of the experimental procedure outlined above are depicted in Fig.~\ref{fig_3} for an initial global phase of $\Phi_0=-1.25$\,rad.
As mentioned earlier, the mean atom number amounts to $\bar{N}=3500$.
Moreover, the thermal initial conditions correspond to a temperature of $T=20$\,nK.

The time-evolution of the local relative phase $\phi(z,t)$ 
of a selected ($N=3500$) single run
is shown in the upper panel of Fig.~\ref{fig_3}\,(a). 
The corresponding time-evolution of the global relative phase $\Phi(t)$ is shown in the lower panel.
In the real experiment it is impossible to observe the time-evolution of the relative phase in a selected single
single run.
Instead the experiment is repeated many times 
until meaningful statistical values can be extracted.
We therefore depict in the upper and lower panel of Fig.~\ref{fig_3}\,(b) the ensemble average of the local relative 
phase $\langle \phi(z,t) \rangle$
and the global relative phase $\langle \Phi(t) \rangle$, respectively.
We would like to recall that $\langle \dots \rangle$ denotes the ensemble average over 
$n_\mathrm{sr} = 301$ independent realizations, where we additionally take into account fluctuations of the total atom number $N$ in accordance with the experiment.
As the atom number difference is the canonical conjugate variable of the phase difference
its time-evolution does not provide any new information and will therefore be omitted for brevity.

The coherent splitting of a single condensate leads to the excitation of a common breathing mode, due to the halving of the atom number within each well. This breathing mode can be easily seen in Fig.~\ref{fig_3} by the pair of black lines marking the region containing $75\%$ of all particles. We note that the observed period $T_\mathrm{b} \approx 48\,\mathrm{ms}$ for $\nu_\parallel = 12\,\mathrm{Hz}$ agrees perfectly with the theoretical prediction $\nu_\mathrm{b} = \sqrt{3} \, \nu_\parallel$ for the breathing mode of a one-dimensional condensate. For the timescales considered, we find the breathing mode to be sufficiently decoupled from the Josephson oscillation dynamics.

\subsection{Local density approximation and dephasing}

A first insight into the evolution of the system can be gained considering the local density approximation (LDA). In the Thomas-Fermi approximation the mean-field density profile is given by an inverted parabola $\rho_0(z) = n_0 (1 - \dimlessZ^2)$. Here $n_0$ is the peak density and we defined the dimensionless spatial coordinate $\dimlessZ = z/R$, where $R$ is the Thomas-Fermi radius. Defining the local chemical potential $\mu(z) = g \rho(z)$ leads to a spatially dependent Josephson frequency 
\begin{align} \label{eq:wJ_LDA}
\omega_{J}(z) = \sqrt{1-\dimlessZ^2} \omega_{J0} ~,
\end{align}
with $\omega_{J0} = \sqrt{4 J n_0 / \hbar}$. The resulting phase profile 
\begin{align}
\phi (z,t) = \Phi_0 \cos \left( \sqrt{1 - \dimlessZ^2} \omega_{J0} t \right)
\end{align} 
describes an assembly of independent undamped Josephson junctions along the weakly confined longitudinal direction. The breathing of the condensate leads to a minor slow time dependence of $\omega_{J0}$, which can be taken into account but does not significantly alter the results. We therefore in the following neglect the influence of the breathing for brevity.

\begin{figure}[b!]
\centering
\includegraphics[width=0.45\textwidth]{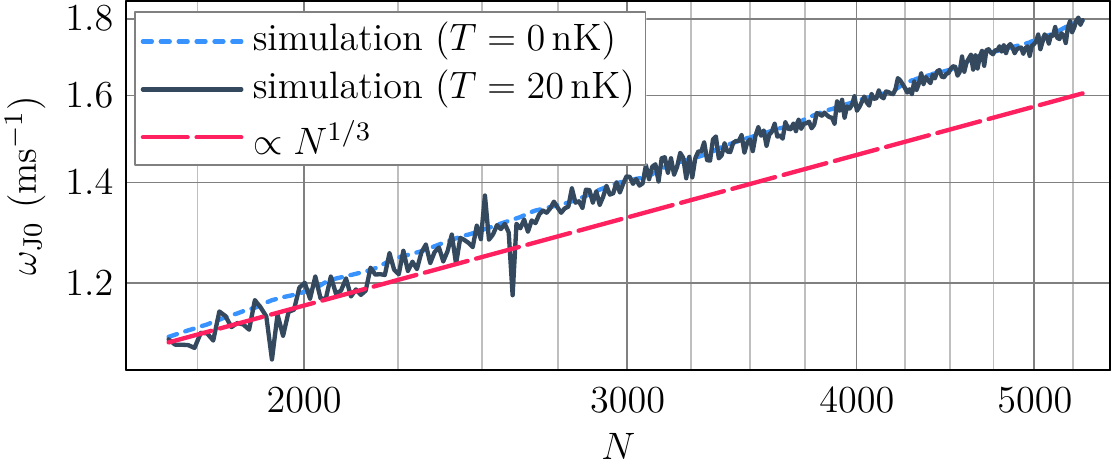}
\caption{
Dependence of the tunneling coupling on the mean field density. The Josephson frequency $\omega_{J0}$ shows additional dependence on the atom number $N$ ($\kappa>1/3$) due to the correction $F$ in Eq.~\eqref{eq:nonLinear_couplingJ}.
}
\label{fig_4}
\end{figure}

The spatial dependence of the Josephson frequency, decreasing towards the edges of the condensate, is clearly visible in Fig.~\ref{fig_3}\,(a) through the bending of the local phase difference. Note that we are not in the self-trapped regime \cite{farolfi2020quantum}. Therefore, different parts of the condensate exhibit local dephasing. This leads to a damping of the global phase difference
\begin{align} \label{eq.struve_single}
\Phi(t) &= \frac{1}{2} \int_{-1}^1 d\dimlessZ\, \Phi_0 \cos \left(  \sqrt{1-\dimlessZ^2} \omega_{J0} t\right) \nonumber \\
&= \Phi_0 \frac \pi 2 \mathbf{H}_{-1} (\omega_{J0} t) ~,
\end{align}
even in the absence of dynamics along the longitudinal direction. Here $\mathbf{H}_{\alpha } (\tau )$ is the Struve function of the order $\alpha $, predicting a power-law decay for the amplitude of the global Josephson oscillation \cite{AbrSteg}.

\subsection{Atom number fluctuations}

The damping of the local and global relative phase in case of the ensemble averages is even stronger. Atom number fluctuations contribute to an additional 
dephasing between different realizations due to the dependence of the 
Josephson frequency $\omega_{J0}$ on the total particle number $N$. The most simple model Eq.~\eqref{eq:wJ_LDA} for the case of a harmonic longitudinal confinement yields $\omega_{J0} \propto N^\kappa$ with $\kappa=1/3$. However, by fitting a whole series of 3D simulations covering a wide range of atom numbers, we find $\kappa \approx 0.43$ (see Fig.~\ref{fig_4}), indicating an interwell tunneling that depends on the local density of the quasicondensates (c.f.~e.g.~\cite{momme2019collective}). Note that we chose the power-law dependence here based on the functional form predicted by Eq.~\eqref{eq:wJ_LDA} and do not expect this relation to hold for notably smaller or larger atom numbers $N$. This also increases the influence of trivial dephasing caused by atom number fluctuations. Eq.~\eqref{eq.struve_single} can be extended to include such fluctuations of the Josephson frequency. Assuming a uniform distribution in $\omega_{J0}$ on the interval $[\bar \omega _{J0} (1-\eta ), \, \bar \omega _{J0} (1+\eta)]$ with $\eta \ll 1$ this leads to
\begin{align} \label{eq.struve_mean}
&\langle \Phi(t) \rangle_N = \frac{1}{2\eta \bar \omega _{J0}}
\int_{\bar \omega _{J0} (1-\eta )}^{\bar \omega _{J0} (1+\eta )}
d\omega\, \Phi _0 \frac \pi 2 \mathbf{H}_{-1} (\omega t) \nonumber \\
&\qquad = 
\Phi _0 \frac \pi 2
\frac{\mathbf{H}_0[ \bar \omega _{J0}t(1+\eta) ]-\mathbf{H}_0 [\bar \omega_{J0} t (1-\eta )]}
{2\eta \bar \omega_{J0} t} ~.
\end{align}
It is worth pointing out that fitting Eq.~\eqref{eq.struve_mean} to the ensemble average $\langle \Phi(t) \rangle$ of our full numerical simulations, we find $\eta$ to be reasonably close (within $\approx 10\%$) to its expected value $\eta = \kappa \, \delta_N/\bar{N}$ with $\kappa \approx 0.43$.

\begin{figure}[t!]
\centering
\includegraphics[width=0.45\textwidth]{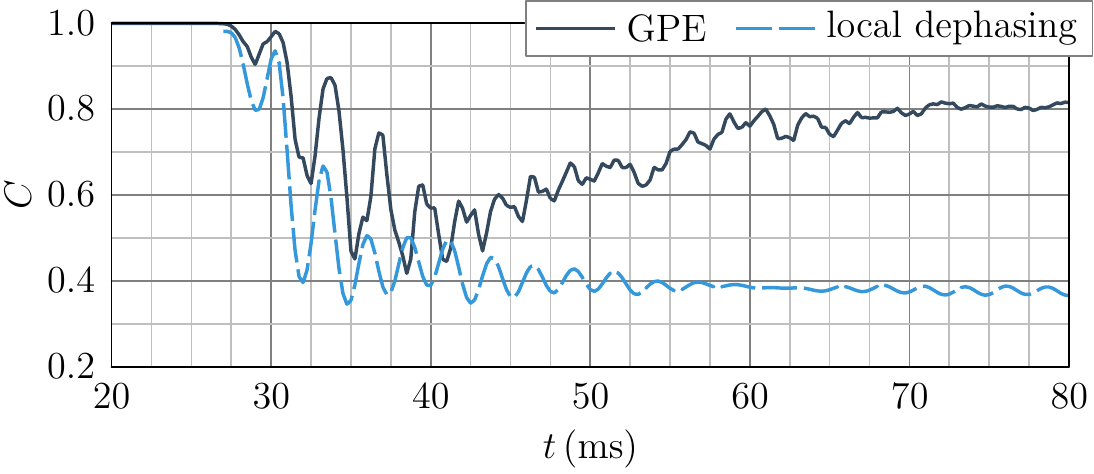}
\caption{ 
Time evolution of the contrast $C(t)$ for the GPE simulation (black line) reveals relaxation beyond local and atom number dephasing (blue dashed line) towards a phase locked state.
}
\label{fig_5}
\end{figure}

\subsection{Relaxation beyond dephasing}

Despite being a good approximation for the evolution of $\langle \Phi(t) \rangle$, these effects alone do not correctly describe the observed relaxation of the system to a phase locked state. Firstly, the small variance of $\langle \Phi(t) \rangle$ at late times (see Fig.~\ref{fig_3}(b)) shows that dephasing due to total atom number fluctuations is not the dominant effect for the damping. Equivalently, this implies damping of the Josephson oscillations within each realization. Secondly, while damping of $\Phi(t)$ is expected from Eq.~\eqref{eq.struve_single}, the observed \emph{local} damping signals the breakdown of the LDA, which would predict an undamped oscillation for fixed position $z$ (c.f.~Fig.~\ref{fig_3}(a)). This difference can be clearly seen in the time evolution of the contrast $C(t)$ depicted in Fig.~\ref{fig_5}, which increases at late times back close to its initial value. This, consistent with \cite{Pigneur18}, is a clear indication for the relaxation to a phase locked state with small fluctuations of the relative phase along the whole condensate. Contrary, considering only local and atom number dephasing $C(t)$ remains small despite the strongly damped global phase oscillations. Note that atom number fluctuations have only a minor influence on the evolution of the contrast $C(t)$, as it is independent of the global phase $\Phi(t)$.

\subsection{Comparison to the experiment}

\begin{figure}[b!]
\centering
\includegraphics[width=0.485\textwidth]{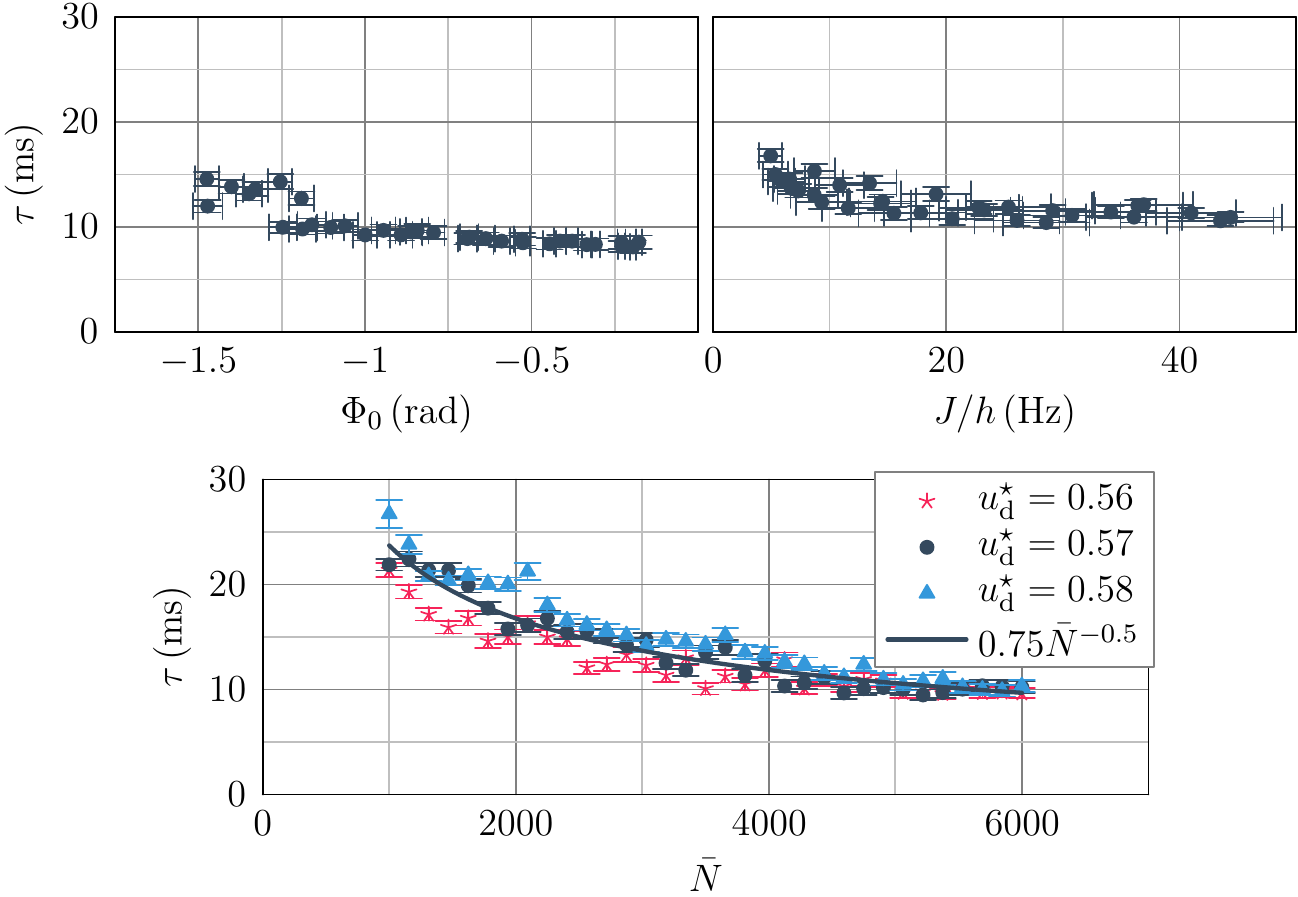}
\caption{
Numerical results of a comprehensive simulation study illustrating the
dependence of the damping time $\tau$ on the initial relative phase $\Phi_0$, 
tunneling strength $J$, and mean atom number $\bar{N}$.
The final dressing amplitude $\potContDistFinal$ determines the 
tunneling coupling (decreasing for larger $\potContDistFinal$). If not explicitly depicted in the sub-figure, the remaining constant parameters are $\bar{N} = 3500$, $\potContDistFinal = 0.56$, and $\Phi_0 = -1.25$, respectively. In all simulations the thermal initial conditions correspond to a temperature of $T = 20$\,nK and the atom number distribution is characterized by the parameters $\sigma_N = 0.16 \bar{N}$ and $\delta_N=0.08 \bar{N}$ (see Sec.~\ref{sec:2_2} for details). Errorbars depict the standard deviation. 
}
\label{fig_6}
\end{figure}

Finally, in a comprehensive simulation study we investigated the dependence of the damping time $\tau$
on the initial relative phase $\Phi_0$, the tunneling strength $J$ and the mean atom number $\bar{N}$ (see Fig.~\ref{fig_6}).
In this context the damping time $\tau$ was determined by fitting 
the results of the numerical simulations using the model 
for a damped Josephson junction presented in \cite{Pigneur18}, where a phenomenological damping term was added to Eq.~\eqref{eq:JJ_density} (see e.g. \cite{Marino99}).
Due to the large number of different parameters it was necessary to reduce the number of
single runs to $n_\mathrm{sr}=21$. 
However, convergence of the extracted values was verified by (selected) computations
using larger values of $n_\mathrm{sr}$. 
In all cases we found our results to be compatible with the experiment~\cite{Pigneur18}. 
This includes a relatively weak dependence of $\tau$ on the initial global relative phase $\Phi_0$, 
a plateau-shaped dependence on the effective tunnel coupling strength $J$, 
and approximately $\tau \sim N^{-0.5}$. 
Note, however, that for the latter we find a slight additional dependence of the 
exponent on the single-particle tunneling coupling $J$.

\section{Multimode dephasing and nonlinear relaxation} \label{sec:4}

In order to study the fundamental mechanisms leading to the rapid local damping of the Josephson oscillations, we consider in the following a zero-temperature ($T=0$) state with fixed particle number $N = 3500$. We further omit the preparation phase to mitigate the effect of a common breathing excitation and imprint the phase difference $\Phi_0$ directly in the final trap configuration. These simplifications preserve the observed main results of rapid global and local damping of Josephson oscillations and enable us to compare results of the full 3D simulations with a tractable linearized, one-dimensional model (see e.g. \cite{momme2019collective,Momme2020}).

\subsection{Effective one-dimensional description} \label{sec:4_1}

We first proceed with the common dimensional reduction, reducing the system to two one-dimensional coupled quantum wires by integrating over the tightly confined transverse directions \cite{olshanii1998atomic,salasnich2002effective}. Regaining the dominant terms in the expansion, the system of two coupled 1D GPE equations can be described by the Hamiltonian
\begin{align}
H_{1\mathrm{D}} = \int dz \, &\sum_{i=\mathrm{L,R}} \left[ \psi_i^\dagger(z) U_\mathrm{GP}\left[\psi_i\right] \psi_i(z) \right ] \nonumber \\
&- \psi_\mathrm{L}^\dagger \mathcal{J}\left[\psi_\mathrm{L},\psi_\mathrm{R}\right] \psi_\mathrm{R} + \mathrm{H.c.} ~.
\end{align}
The uncoupled evolution of the fields $\psi_\mathrm{L,R}$ in the left (right) well of the double-well potential is described by
\begin{align}
U_\mathrm{GP}\left[\psi_i\right] = -\frac{\hbar^2}{2 m} \partial_z^2 + V(z) + \frac{g_{1\mathrm{D}}}{2} |\psi_i(z)|^2 ~,
\end{align}
where $g_{1\mathrm{D}} \approx 2 \hbar a_s \omega_\perp$ is the effective 1D interaction constant. The tunneling coupling is given by
\begin{align} \label{eq:nonLinear_couplingJ}
\mathcal{J}\left[\psi_\mathrm{L},\psi_\mathrm{R}\right] = \hbar J + \frac{\hbar F}{2} \left[ |\psi_\mathrm{L}|^2 + |\psi_\mathrm{R}|^2 \right] ~,
\end{align}
where we included the dominant nonlinear density dependence (see e.g.~\cite{momme2019collective} for details). In general we have $|J| \gg |F \operatorname{max}[\rho_0]|$. Note, in particular, that also $J$ can show an explicit density dependence due to radial swelling of the condensate \cite{salasnich2002effective}. We find the dominant effect of the nonlinear density dependence to be a moderate shift of $\omega_{J0}$ depending on the total atom number $N$ (see Fig.~\ref{fig_4}). We therefore neglect in the following linearized model the nonlinear terms in Eq.~\eqref{eq:nonLinear_couplingJ}, i.e.~$F=0$, and treat $J$ as a free parameter to be determined from our 3D simulations.

Next, writing the fields in the Madelung representation $\psi_\mathrm{L,R} = \sqrt{\rho_0 + \delta\rho_\mathrm{L,R}}\operatorname{e}^{i \phi_\mathrm{L,R}}$, we expand in powers of the small density perturbations $\delta \rho$ and phase gradients $|\partial_z \phi|$ (where $|\dots|$ denotes the typical value) \cite{mora2003extension}. The Hamiltonian to quadratic order separates into a weakly coupled sum $H \approx H_s + H_a$ for the symmetric (\textit{s}) and antisymmetric (\textit{a}) degrees of freedom (DoF), defined as
\begin{align}
\phi_s = \frac{1}{2} \left( \phi_1 + \phi_2 \right) &\qquad \phi_a = \phi_1 - \phi_2 \\
\delta \rho_s = \delta \rho_1 + \delta \rho_2 &\qquad \delta \rho_a = \frac{1}{2} \left(\delta \rho_1 - \delta \rho_2\right) ~.
\end{align}
The Hamiltonians $H_{s,a}$ are given by
\begin{align} \label{eq:H_full_decoupled}
H_{i} &= \! \int \! dz \, \Big\{ - \! \frac{\hbar^2}{4 m} \! \Big[ \zeta_{i}^2 \left(\partial_z \delta \rho_{i}\right)^2 \! + \! \frac{\rho_0^2}{\zeta_{i}^2} \left( \partial_z \phi_i \right)^2 \Big] + \zeta_{i}^2 g_{1\mathrm{D}} \delta \rho_i^2 \nonumber \\
&- \delta_{ia} \Big[ \hbar J \rho_0 \big( \operatorname{cos}(\phi_a) - 1 \big) + \frac{\hbar J \delta \rho_a^2}{2 \rho_0} \operatorname{cos}(\phi_a) \Big] \Big\} \, ,
\end{align}
where $i=s,a$ and we suppressed the spatial and temporal dependence of the fields $\delta \rho(z,t)$, $\phi(z,t)$, and $\rho_0(z)$ for simplicity.
Here $\zeta_{s(a)} = 0.5 \, (1)$ for the symmetric (antisymmetric) DoF, $\delta_{ia}$ is the Kronecker delta, and we included the quantum pressure and minor coupling correction (first and last term respectively) for completeness. If neglected, the symmetric and antisymmetric DoF are described by the Luttinger-Liquid ($H_\mathrm{LL}$) and sine-Gordon ($H_\mathrm{sG}$) model, respectively \cite{Gritsev07}. Note, in particular, that while second order in the small parameters $\delta \rho$ and $|\partial_z \phi_a|$ the sine-Gordon model is an interacting (quantum) field theory with $H_\mathrm{sG}$ including terms with an arbitrary (even) number of fields $\phi_a$. 

\subsection{Analytic solutions for harmonically trapped systems}

Within the framework of the sine-Gordon model the relaxation of Josephson oscillations in a homogeneous system, $\rho_0 = \text{const.}$, was studied in \cite{Foini17,van2019self,horvath2019nonequilibrium}. This was, however, insufficient to describe the observed fast damping in the experiment \cite{Pigneur18}. Here, to investigate the influence of the longitudinal confinement, we present an analytic solution to the linearized equations of motion for a harmonically trapped system.

Under the assumption of (i) the Thomas-Fermi approximation $(\partial_z \rho_0) / \rho_0 \ll 1$, (ii) neglecting the nonlinear density dependence of the tunneling coupling, and (iii) small tunneling energy compared to the chemical potential $\hbar J / \mu \ll 1$ the linearized equations of motion for the relative phase field obey the eigenvalue equation
\begin{align} \label{eq.phase_linear_EoM}
\partial_\dimlessZ \left[ (1-\dimlessZ^2) \, \partial_\dimlessZ \phi \right] + \left[ \lambda_{0n}^2 - \gamma^2 (1-\dimlessZ^2) \right] \phi = 0 ~,
\end{align}
where we assumed $\phi(t) \sim \operatorname{e}^{i\omega_n t}$ and defined the dimensionless eigenvalue $\lambda_{0n} = \sqrt{2} \omega_n / \omega_\parallel$ and  Josephson frequency $\gamma = \sqrt{2} \omega_{J0} / \omega_\parallel$. We recall the definition of the dimensionless spatial coordinate $\dimlessZ = z/R$, where $R$ is the Thomas-Fermi radius. The dominant correction beyond the LDA is given by the kinetic energy, first term in Eq.~\eqref{eq.phase_linear_EoM}. If neglected, we recover the previous model of an assembly of undamped, uncoupled Josephson junctions with spatially dependent frequency. The kinetic energy term couples these oscillators, damping the rapid growth of the phase gradient caused by local dephasing. 

\begin{figure}[t!]
\centering
\includegraphics[width=0.48\textwidth]{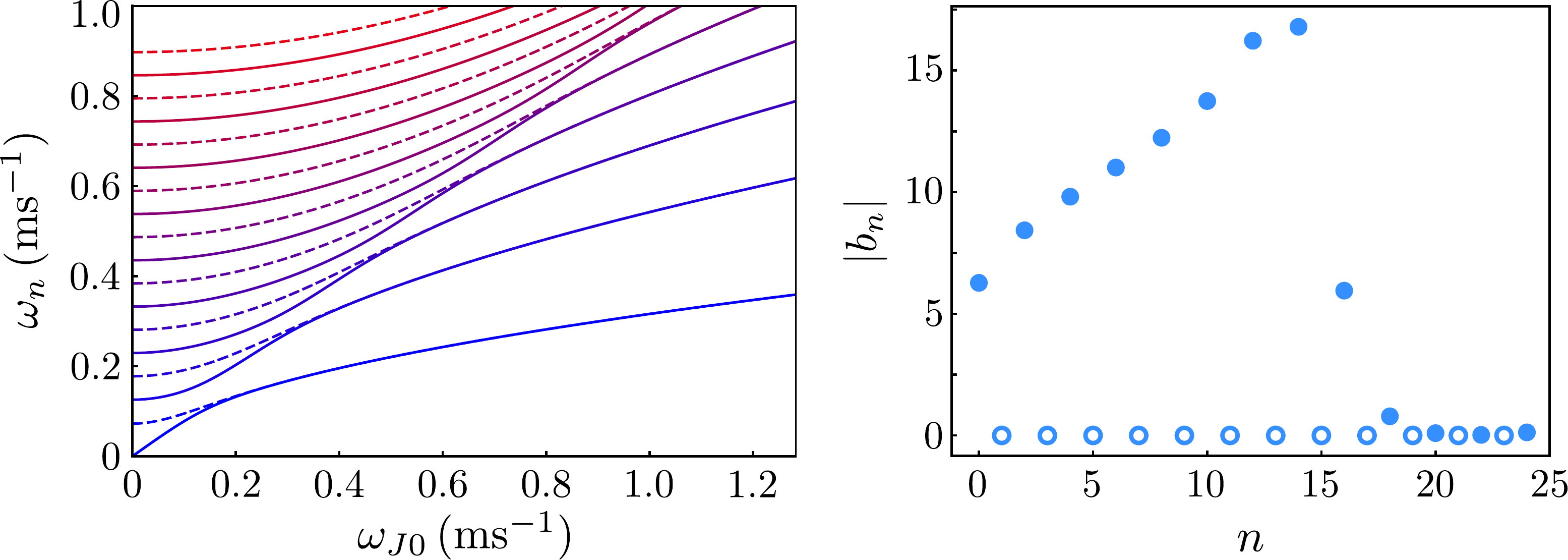}
\caption{
Left panel: Frequencies $\omega_n$ for the first $16$ eigenmodes of the linearized model Eq.~\eqref{eq.phase_linear_EoM} as a function of the tunneling coupling $\omega_{J0}$. Even (odd) modes are depicted with solid (dashed) lines. Right panel: Quasiparticle occupations $|b_n|$ for an initial global phase difference $\Phi_0$. Due to symmetry only even modes (solid markers) are occupied, their occupation decreasing rapidly for $n \geq 15$.
}
\label{fig_7}
\end{figure}

Exact solutions to Eq.~\eqref{eq.phase_linear_EoM} are given by the angular oblate spheroidal wave functions $S_{mn}$ for $m=0$ \cite{AbrSteg}. The eigenfrequencies $\omega_n$ are depicted in the left panel of Fig.~\ref{fig_7}, featuring an increasing energy gap and twofold degeneracy of the spectrum for larger tunneling coupling $J$ (cf. \cite{momme2019collective}). For vanishing tunneling coupling, $\gamma=0$, Eq.~\eqref{eq.phase_linear_EoM} reduces to the Legendre differential equation, leading to the known spectrum for the excitations of the inhomogeneous Luttinger-Liquid model for a harmonically trapped quasi-condensate \cite{petrov2000regimes}. Imposing canonical commutation relations for $\delta\rho_a$ and $\phi_a$ determines the normalization of the mode functions and leads to the mode expansion of the phase and density quadratures
\begin{align}
\phi_a &= \sum_{n} \sqrt{\frac{(2n+1) \mu}{2 R_\mathrm{TF} \hbar \omega_n \rho_0}} S_{0n}(\dimlessZ) \, b_n  \operatorname{e}^{i \omega_n t} + \mathrm{H.c.} \label{eq.mode_expansion_phase} \\
\delta\rho_a &= \sum_{n} \sqrt{\frac{(2n+1) \hbar \omega_n \rho_0}{2 R_\mathrm{TF} \mu}} S_{0n}(\dimlessZ) \, b_n  \operatorname{e}^{i \omega_n t} + \mathrm{H.c.} \label{eq.mode_expansion_density}
\end{align}
In Fig.~\ref{fig_7} (right panel) we show the occupation numbers $|b_n|$, calculated for an initial constant phase difference $\phi(z) \equiv \Phi_0$ by projecting onto the quasiparticle basis, Eq.~\eqref{eq.mode_expansion_phase}. In contrast to the spatially homogeneous system, where a constant global phase difference $\Phi_0$ only has non-vanishing overlap with a single ($k=0$) mode, here multiple modes are populated. Due to symmetry only even modes are occupied. Their amplitude decreases rapidly for $n>15$, such that only the 10 lowest energy modes show significant population. 

\begin{figure}[b!]
\centering
\includegraphics[width=0.48\textwidth]{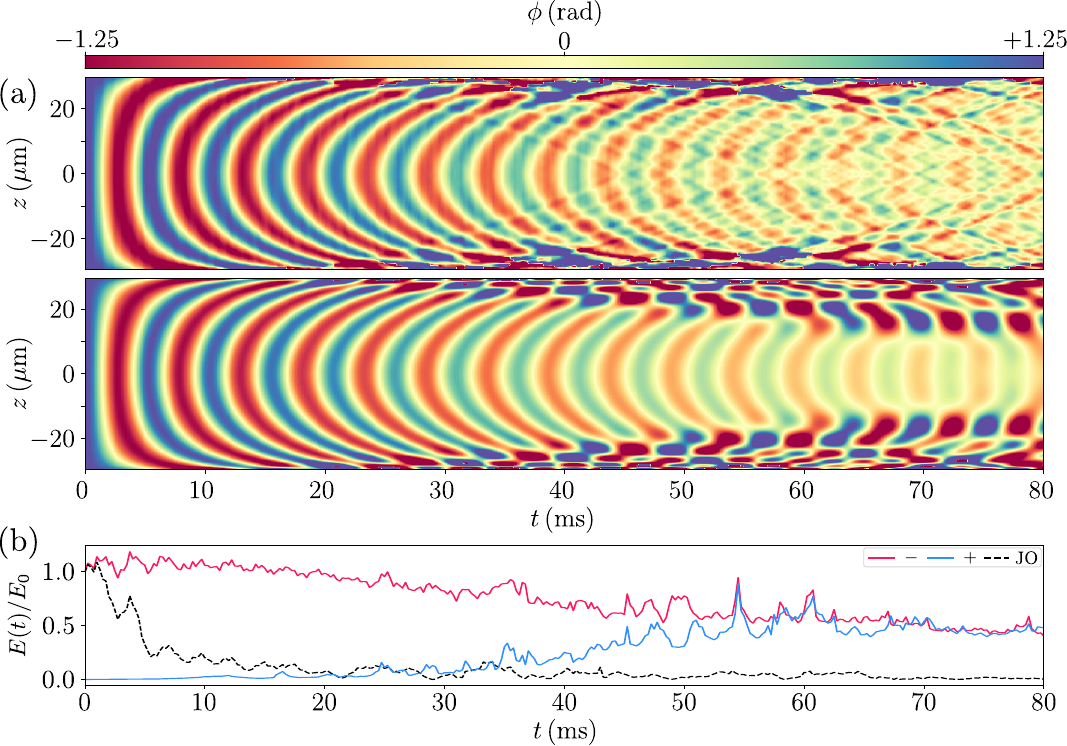}
\caption{
(a) Comparison of the phase difference $\phi(z,t)$ from the full 3D-GPE simulations (upper panel) to the 1D model Eq.~\eqref{eq.mode_expansion_phase} (middle panel). The system parameters are $N=3500$, $T=0 \, \mathrm{nK}$, $\Phi_0 = -1.25$\,rad, and $\omega_{J0} \approx 235 \, \mathrm{s}^{-1}$. In (b), the normalized energy in the symmetric (blue line) and antisymmetric (red line) degrees of freedom and the $k=0$ mode (black dashed line) reveal multimode dynamics and relaxation beyond the SG model (see main text for details). Note that, while the GPE obeys energy conservation, the depicted energies are calculated using Eq.~\eqref{eq:H_full_decoupled} and hence are not strictly conserved.
}
\label{fig_8}
\end{figure}

We compare in Fig.~\ref{fig_8}(a) predictions of our analytic model to the evolution of the local relative phase $\phi(z,t)$ obtained from the full 3D GPE simulation. As previously mentioned, we consider a fixed particle number $N=3500$, zero temperature $T=0$, and omit the preparation phase to suppress the common breathing mode. While, e.g., the damping time $\tau$ depends on details of the initial state, these simplifications qualitatively preserve the main features of the relaxation (c.f.~Fig.~\ref{fig_3}). In particular, the system still exhibits fast local damping of the Josephson oscillation and relaxes to a phase locked state on long timescales. 

The time evolution in the harmonic approximation is given by the undamped oscillations between density and phase quadratures of the initially populated free quasiparticle modes (Eq.~\eqref{eq.mode_expansion_phase}). Each mode oscillates with its distinct eigenfrequency $\omega_n = \lambda_{0n} \omega_\parallel / \sqrt{2}$, where we treated the Josephson frequency $\omega_{J0}$ as a free parameter in Eq.~\eqref{eq.phase_linear_EoM}, determined from the GPE simulations. The resulting multimode dephasing leads to a rapid initial damping of the global Josephson oscillation, similar to the previous local dephasing. Most notably, however, the dephasing of free quasiparticle modes already leads to a local damping of oscillations. At early times, we find excellent agreement between the full numerical simulation and the analytic predictions. For later times the linearized theory shows oscillations increasing again in amplitude, which propagate inwards from the boundaries. These are caused by (partial) rephasing dynamics as a result of the limited number of quasiparticle modes with significant population \cite{rauer2018recurrences}. 

The absence of rephasing in the full 3D simulations signals the breakdown of the linearized model, either through nonlinear terms in the SG model or coupling of the symmetric and antisymmetric DoF. For the former the Hamiltonian of the system is still given by $H \approx H_s + H_a$, such that the symmetric DoF can be neglected. Higher-order corrections couple the symmetric and antisymmetric DoF leading to a transfer of energy between the two sectors, ultimately leading to complete equilibration of the system. This signals a definite breakdown for the effective description of the extended Josephson junction through the sine-Gordon model. 

We quantify the coupling in Fig.~\ref{fig_8}(b), showing the time evolution of the energies $E_i(t)$ within each sector $H_{s,a}$ given by Eq.~\eqref{eq:H_full_decoupled} at each instant of time. In addition, we display the energy of the zero-momentum ($k=0$) mode of the antisymmetric DoF determining the spatially independent, global Josephson oscillation. Therefore, the initial energy $E_0$ is completely given by this zero-momentum mode. At early times the energies are approximately conserved within each sector, validating a description of the antisymmetric DoF in terms of the sine-Gordon model. For the parameters considered the linearized model (Eq.~\eqref{eq.phase_linear_EoM}) constitutes a good approximation in this regime, signaling that nonlinearities of the sine-Gordon model only lead to minor corrections. The rapid decline of the $k=0$ mode is caused by the dephasing of quasiparticle modes, transferring energy to higher momentum states within the antisymmetric sector. Note that quasiparticle dephasing here leads to transport in momentum space since plane waves are not the eigenstates of the system.

For $30\,\mathrm{ms} \lesssim t \lesssim 60\,\mathrm{ms}$ energy transfer from the antisymmetric to the symmetric DoF becomes dominant, revealing the breakdown of the SG model. At later times the system reaches equipartition of energy. This does however not imply complete thermalization of the system. We like to highlight that significant coupling already occurs at zero temperature due to the spatially inhomogeneous Josephson frequency. This is the dominant effect leading to the fast relaxation of the system to a phase locked state as observed in \cite{Pigneur18}.

\begin{figure}
\centering
\includegraphics[width=0.48\textwidth]{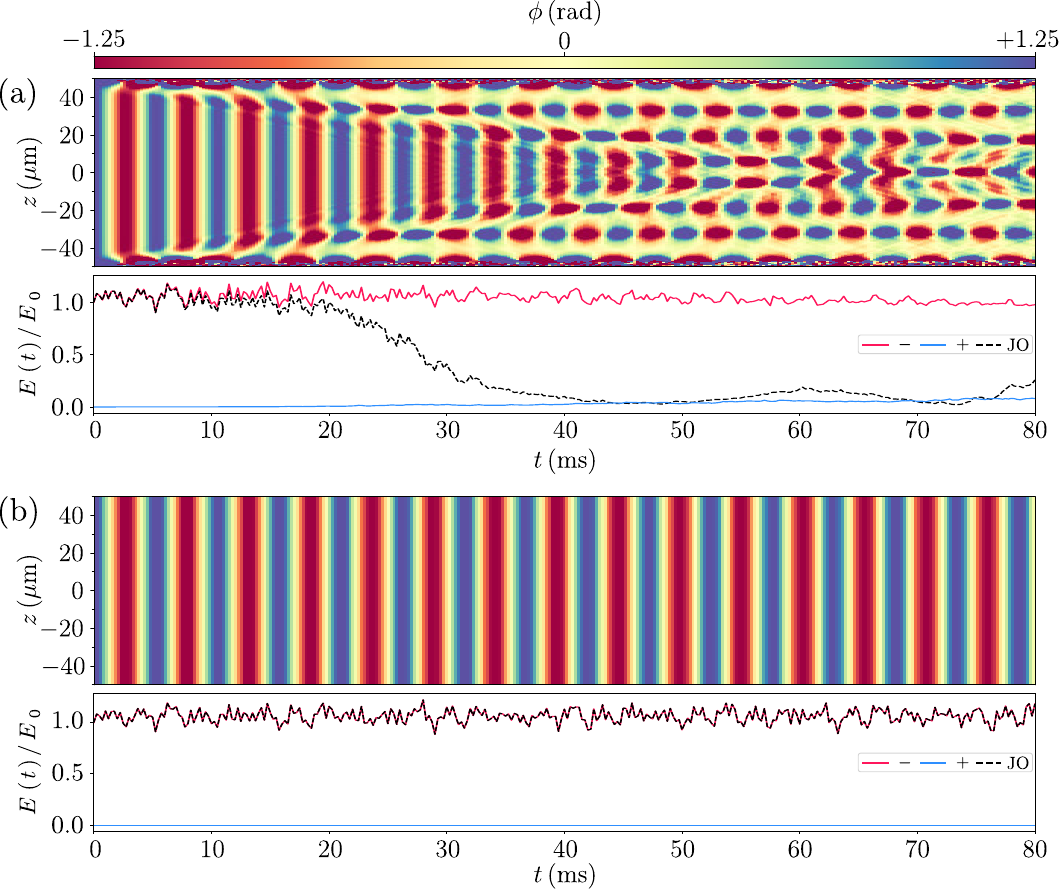}
\caption{ Relaxation in flat-bottom potentials. Results of 3D-GPE simulations for parameters as in Fig.~\ref{fig_8} for a box-shaped potential (a) and a ring-shaped potential (b) in the longitudinal $z$-direction. We adjusted the particle number $N$ requiring equal Josephson frequencies $\omega_J(z\!=\!0)$ for all trapping potentials. In neither case the system relaxes to a phase locked state since coupling of symmetric and antisymmetric degrees of freedom is highly suppressed. For (b) the Josephson oscillation completely decouples.
}
\label{fig_9}
\end{figure}

\subsection{Flat-bottom potentials}

The rapid equipartition of energy greatly impedes experimental studies of, e.g., the influence of quantum corrections or the long time evolution of extended bosonic Josephson junctions and the sine-Gordon model. Based on recent progress in shaping arbitrary trapping potentials \cite{Tajik19,Heathcote08} we now give an outlook on the possibilities for mitigating these effects in cold-atom experiments. 

We show in Fig.~\ref{fig_9}(a) the time evolution of $\phi(z,t)$ and the normalized energies for a box-shaped potential, which is known to violate the system integrability, even if the potential is flat between the walls \cite{Brezinova2011}. In accordance with typical experimental capabilities, we model the $z$-dependence of $V_\mathrm{rf}$ in Eq.~\eqref{eq:V} as
\begin{align}
V(z) = \frac{\hbar V_0}{2} \Big[ \tanh \Big( \frac{|z| - L/2}{\sigma_w} \Big) + 1 \Big] ~,
\end{align}
with $V_0 \approx 15 \, \mathrm{kHz} \gg \mu$, length $L=100 \, \mu\mathrm{m}$, and finite wall width $\sigma_w = 2 \, \mu\mathrm{m}$. Spatial dependence of the Josephson frequency is limited to the edges of the condensate leading to disturbances emanating from the boundaries. Once propagated inwards these disturbances lead to a rapid decline of the $k=0$ mode caused by multimode dephasing. In contrast to the harmonic confinement, however, global Josephson oscillations prevail within the central region of the box (for $t \lesssim 20\,\mathrm{ms}$). In particular, the amplitude of local phase oscillations remains high. Therefore, the contrast $C(t)$ after its initial decrease remains small at longer times, i.e.~the system does not relax to a phase locked state. Consistently, coupling between the symmetric and antsymmetric DoF is highly suppressed, with only $\approx 16\%$ of the energy being transferred at $t=80\,\mathrm{ms}$. Therefore, while the antisymmetric DoF shows multimode dynamics, decoupling from the symmetric DoF constitutes a good approximation.

Ring-shaped potentials further eliminate the influence of the boundaries, leading to undamped global Josephson oscillations at zero temperature (see Fig.~\ref{fig_9}(b)). In accordance, we find the total energy to remain in the $k=0$ mode, with negligible coupling between the symmetric and antisymmetric DoF. Here, dephasing due to atom number fluctuations will dominate ensemble averages at late times, which can be mitigated through appropriate postselection. Naturally, this is the ideal setting to study the nonlinear dynamics and the influence of thermal and/or quantum fluctuations. 

\section{Conclusion} \label{sec:5}

We gave a detailed discussion of the rich nonlinear dynamics in inhomogeneous extended bosonic Josephson junctions. We found our results for the full 3D-GPE simulations at finite temperature to reproduce the experimental findings \cite{Pigneur18} over a wide range of parameters. 

A detailed analysis for the zero temperature case allowed us to distinguish two stages of the relaxation dynamics. The short time behavior was well described through the sine-Gordon model, i.e.~the low-energy effective theory for the antisymmetric degrees of freedom of two tunnel-coupled one-dimensional superfluids. For the parameters considered, we found reasonable agreement with an analytic solution in the harmonic approximation. In contrast to the local density approximation, these solutions already predict the local damping of Josephson oscillations (in addition to the spatially dependent Josephson frequency). At later times, we explained the relaxation to a phase locked state, as observed in \cite{Pigneur18}, through the breakdown of the sine-Gordon model description by coupling of the symmetric and antisymmetric degrees of freedom. Induced by the harmonic confinement, this coupling dominates the long time behavior of the system, even at zero temperature. Lastly, we showed that this coupling can be greatly reduced for box- or ring-shaped potentials along the longitudinal direction. 

Our study is a crucial step when investigating the influence of thermal or quantum fluctuations on the relaxation dynamics in bosonic Josephson junctions. Our simulations provide, e.g., the relevant maximum timescale during which comparison to the sine-Gordon model is sensible. A detailed comparison to other microscopic models, such as the self-consistent time-dependent Hartree-Fock approximation presented in \cite{van2020josephson}, would be interesting but is hindered by the different regimes of applicability. Therein, results are presented in the small particle number regime $N \lesssim 200$, wherein deviations of classical statistical simulations are expected to become relevant. Intriguingly, the behavior in \cite{van2020josephson} is similar to our harmonic 1D model, including local damping and recurrences of phase oscillations for harmonic confinements, and the absence of relaxation to a phase locked state. This, consistent with the considered small atom number limit, suggests that interactions play a less significant role in the regime of Ref.~\cite{van2020josephson}.

Our results highlight the importance of understanding the classical nonlinear dynamics, even at zero temperature, before comparing results to sophisticated quantum models. Our work outlines the experimental possibilities and steps necessary to study relaxation in extended bosonic Josephson junctions and the (quantum) sine-Gordon model out of equilibrium.

\vspace{1cm}
We acknowledge support by the Wiener Wissenschafts- und TechnologieFonds (WWTF), project No. MA16-066 (``SEQUEX'') and the Austrian Science Fund (FWF) via Grant SFB F65 (Complexity in PDE systems) and SFB 1225 Project-ID 273811115 (ISOQUANT). S.E.~acknowledges support from the European Union’s Horizon 2020 research and innovation programme under the Marie Sklodowska-Curie grant agreement No 801110 and the Austrian Federal Ministry of Education, Science and Research (BMBWF) from an ESQ fellowship.

\bibliographystyle{apsrev4-1}

%

\end{document}